# Extracting Particle Size Distribution from Laser Speckle with A Physics-Enhanced AutoCorrelation-based Estimator (PEACE)

**Author list**

Qihang Zhang[1], Janaka C. Gamekkanda[2], Ajinkya Pandit[2], Wenlong Tang[3], Charles Papageorgiou[4], Chris Mitchell[4], Yihui Yang[4], Michael Schwaerzler[5], Tolutola Oyetunde[5], Richard D. Braatz[2], Allan S. Myerson[2], and George Barbastathis[6,7]*

1. Department of Electrical Engineering and Computer Science, Massachusetts Institute of Technology, Cambridge, Massachusetts 02139, USA.

2. Department of Chemical Engineering, Massachusetts Institute of Technology, Cambridge, Massachusetts 02139, USA.

3. Data Sciences Institutes, Takeda Pharmaceuticals International Co, 650 E Kendall St, Cambridge, Massachusetts 02142, USA.

4. Process Chemistry Development, Takeda Pharmaceuticals International Co, 40 Landsdowne St, Cambridge, Massachusetts 02139, USA.

5. Innovation and Technology Sciences, Takeda Pharmaceutical Company Limited, 200 Shire Way, Lexington, MA 02421, USA.

6. Department of Mechanical Engineering, Massachusetts Institute of Technology, Cambridge, Massachusetts 02139, USA.

7. Singapore-MIT Alliance for Research and Technology (SMART) Centre, 1 Create Way, Singapore 117543, Singapore.

* Email: gbarb@mit.edu

**Abstract**

**Extracting quantitative information about highly scattering surfaces from an imaging system is challenging because the phase of the scattered light undergoes multiple folds upon propagation, resulting in complex speckle patterns. One specific application is the drying of wet powders in the pharmaceutical industry, where quantifying the particle size distribution (PSD) is of particular interest. A non-invasive and real-time monitoring probe in the drying process is required, but there is no suitable candidate for this purpose. In this report, we develop a theoretical relationship from the PSD to the speckle image and describe a physics-enhanced autocorrelation-based estimator (PEACE) machine learning algorithm for speckle analysis to measure the PSD of a powder surface. This method solves both the forward and inverse problems together and enjoys increased interpretability, since the machine learning approximator is regularized by the physical law.**

Speckle results from the propagation of a wavefront whose phase has been strongly modulated by spatially variant features across a surface (or volume), so the speckle is an encoding of spatial patterns on the rough surface. As long as the morphology statistics are invariant, it is straightforward to relate statistical moments of the surface to the statistical moments of the speckle[1–3]. The laser speckle pattern has long been used to characterize surface roughness[3–12]. However, the method typically works only when the surface height fluctuation (equivalently, typical particle size) is smaller than or comparable to the light wavelength[3], limiting its application to surfaces encountered in many industrial processes, such as pharmaceuticals manufacturing. Electronic Speckle Pattern Interferometry (ESPI) can measure the surface motion distribution even at nanometer scales, but it is a two-step measurement with a reference beam, inhibiting real-time monitoring and lacking absolute height information[13–15]. Laser speckle contrast imaging is qualitative and does not yield quantitative surface roughness[16]. Interferometric particle imaging can measure the particle size and shape, but it only works for a single particle or sparsely distributed particles[17–19].

Recent advances in Machine Learning have been successful in imaging through scattering media[20–25] and speckle suppression[26–31]. However, in both cases, the speckle pattern is treated as an unwanted disturbance. Extracting the scattering media information from the speckle has also been pursued qualitatively: for example, the classification of materials according to the scattered speckle patterns[5–8]. The main difficulty preventing further quantitative speckle analysis is the sensitivity of the phase signal to surface randomness, which hinders neural networks' ability to identify other underlying dynamics.

Quantitative granularity characterization is desired in many applications[32–34], particularly the powder drying process in the pharmaceutical industry, during which the wet solid ("cake") is converted into a powder consisting of particles with the requisite size distribution. These powders are subsequently employed with other ingredients to form solid oral dosage forms such as tablets and capsules. However, agglomeration, deagglomeration and crystal breakage are all likely during the drying process. Occurrence of hard agglomerates could influence content uniformity and functionality in the final drug product, e.g., if the active ingredient concentration becomes too high. Even though the parameters of the drying processes are generally well-controlled, the evolution of the particle sizes during agitation is not fully predictable. Thus, it is crucial to monitor particle sizes quantitatively in real-time and correct for abnormal size changes through feedback control on process parameters (e.g., temperature, agitation speed).

No real-time online monitoring methods exist presently, to our knowledge, that can detect early on and prevent such abnormal particle size changes for wet powder drying. Since the cake surfaces closely meet the Lambertian assumption[35,36], imaging by a standard camera from a distance compatible with the manufacturing setting (~0.2 – 0.5 m away from the powder) does not provide sufficient contrast and spatial resolution to extract the surface PSD. Machine vision to analyze the appearance of the cake surface and detect agglomerates is generally limited due to the same reason[37]. Imaging with *in situ* cameras relies on particle sparsity, which only works for solid suspensions rather than wet powder[38–40]. Moreover, this method is invasive, so there is a risk of the powder obscuring the viewing field and rendering the imaging operation impossible. Instead, manufacturers commonly rely on trained personnel to visually observe the mixing—but this can be subjective. Lastly, it is possible to extract a sample from the cake at fixed time intervals and pass it through a particle size analyzer instrument. However, this method is invasive and slow and, thus, not suitable for industrial use.

In this work, we propose a physics-enhanced autocorrelation-based estimator (PEACE) to extract the PSD of a powder surface from its laser speckle as shown in Fig. 1. With the help of the free-space propagation equations, we relate the ensemble-averaged spatial-integral autocorrelation function to the statistics of powder surface, *i.e.,* the PSD. This relationship becomes the forward model for the estimator, yet it is inevitably incomplete. For example, particles may overlap along the longitudinal direction, which should not be introduced in the explicit model lest it becomes exceedingly complicated. Similarly, our experimental approach includes a finite spatial integral and temporal averaging of several frames, which are subject to sensor uncertainties in the model. Another limitation is that collecting sufficient experimental data to compensate for these uncertainties is prohibitively expensive.

We use PEACE to compensate for these uncertainties in the forward model. As shown in Fig. 1b, a small neural network called "generator" is combined with the forward model to form the final map from the PSDs to the experimentally collected autocorrelation images. In this way, starting from a modest amount of experimental data we can create a much larger synthetic (simulated) dataset. Finally, a deep neural network (DNN) called "estimator" is trained by the synthetic dataset to learn the inverse mapping from the speckle autocorrelation to the PSD. The estimator is overparameterized but generalizes well, confirming recent theoretical developments[41]. This method solves both the forward and inverse problems together, improving the machine learning model's generalization ability and interpretability. For example, the explicit forward model allows us to estimate bounds on our prediction ability. We note that the terms "generator" and "estimator" may be reminiscent of the generator and discriminator in Generative Adversarial Networks (GANs)[42] yet our approach is significantly different in that we do not employ adversarial training.

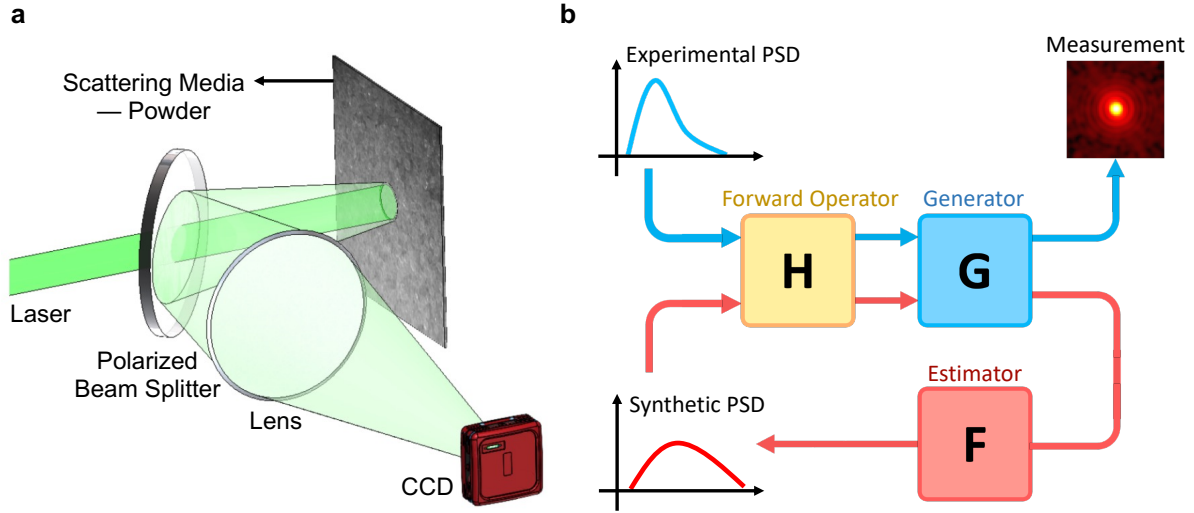

**Fig. 1 | Overview of the speckle probe and the PEACE algorithm. a** A sketch of our speckle probe. We collect the scattered light from the scattering media – powder in our case – with a monochromatic CCD camera. The Machine Learning-based analysis can extract the quantitative surface information, the PSD, from the speckle statistics. **b** PEACE training loop. The forward operator comes from the physics model. A small "generator" with a physics-picture-inspired structure is trained by a modest amount of the experimental data. The forward operator and the trained generator produce a much larger synthetic dataset. This synthetic dataset trains the DNN "estimator" to learn the mapping from the measured speckle autocorrelation to the particle size distribution (PSD). The 'generator' only contains 2.8k parameters, while the estimator has 377k parameters.

Here, we show that our method overcomes these limitations by providing a real-time, non-invasive, far-field optical probe (as shown in Fig. 1a) to monitor particle size distributions quantitatively. Especially for densely concentrated wet powders, this method is the first in-line measurement, and it is easily deployable in the industrial instrument.

## Results

### Forward model – Speckle and Particle Statistics

With the help of the physics model, we derive the expression of the forward operator $H$ in the Method Section and the Supplementary Section 3.

$$\langle A(u) \rangle = H(p(r)) = \frac{1}{C} \frac{4 \sin^2\left(\frac{Du}{2}\right)}{D^2 u^2} \left| \int p(r) \frac{\sin(ru)}{u} dr \right|^2 \tag{1}$$

where $\langle A \rangle$ denotes the ensemble average autocorrelation of the speckle pattern, $p(r)$ is the PSD in terms of number of particles, $D$ is the beam spot diameter which is much larger than the particle size $r$ in our range of interest (50 μm - 1000 μm), and $C = |\int p(r) r dr|^2$ is the normalization factor. Therefore, the main ring-shape feature is dominated by $\frac{4\sin^2\left(\frac{Du}{2}\right)}{D^2 u^2}$, and the side-lobe intensities are modulated by $\left|\int p(r) \frac{\sin(ru)}{u} dr\right|^2$. We design the progression of a typical experiment based on this forward operator as shown in Fig. 2.

Fig. 2a shows two sample sets with different particle sizes imaged through a regular microscope. Fig. 2b are photos collected by a commercial camera, with broadband and spatially incoherent illumination. The contrast and resolution are too coarse to resolve the particles, not displaying any discernible features that could be attributed to particle size. Fig. 2c shows their raw speckle images collected from the CCD camera. Fig. 2d shows the respective speckle autocorrelation patterns, which are subsequently averaged as shown in Fig. 2e. The calculation results in Fig. 2e are passed to a two-stage neural network consisting of the generator and the estimator. The generator corrects the physical model for uncertainties, such as overlapping particles that would be very hard to model analytically. The estimator receives its input from either the generator or the measured results (shown in Fig. 2e) and produces the cumulative distribution of the particle sizes. By one more differentiation step, we obtain the final estimate of the PSD as a vector of powder concentrations *vs.* particle size. Fig. 2f shows one typical test result from a particle distribution never shown during the neural network's training stage, as computed by the combined generator-estimator algorithm. This estimated PSD can be compared with the ground truth PSD that we established using a commercial particle size analyzer, the "Mastersizer" (details in Supplementary Section 2). The complete details of this process are in the Supplementary Sections 3 and 4. Fig. 2g shows cross-sections of the averaged autocorrelation for different sample particle sizes. The intensities of the first and second-order lobes monotonically decrease with ascending PSDs. The higher-order lobes are merged into the background fluctuations for the large particle size sample, while for the small particle size sample the lobes are clearly resolvable.

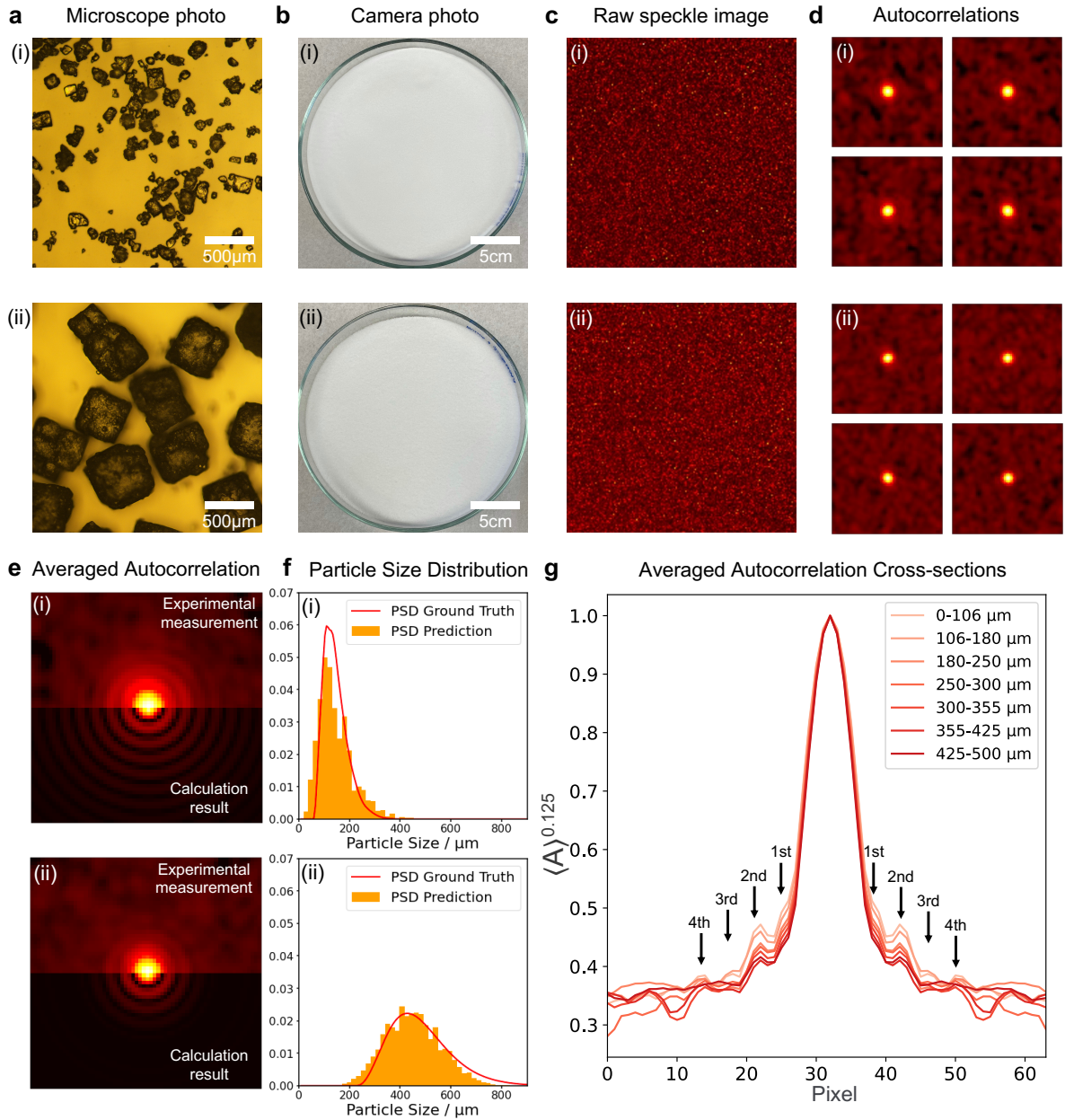

**Fig. 2 | KCl powder results with sizes (i) ~106–180 μm and (ii) ~425–500 μm. a** Microscope and **b** commercial camera photos for two samples. **c** Raw speckle images collected by the CCD camera. **d** These four spatially integrated autocorrelation images are collected on the same sample with different particle positions. **e** The upper half image shows the averaged 1000 autocorrelation frames from the measurement. Since the particle position is ergodic, the temporal average equals the ensemble average of the autocorrelation image. The lower half image shows the calculation result from the forward operator. **f** The ground truth particle size distribution (PSD) and the corresponding estimator prediction are plotted together. **g** Cross-section plots of the averaged autocorrelations in **e** for different samples with ascending PSDs. The positions of the high-order lobes are marked. We raise the autocorrelation to the power of one-eighth to enhance the sidelobes' visibility.

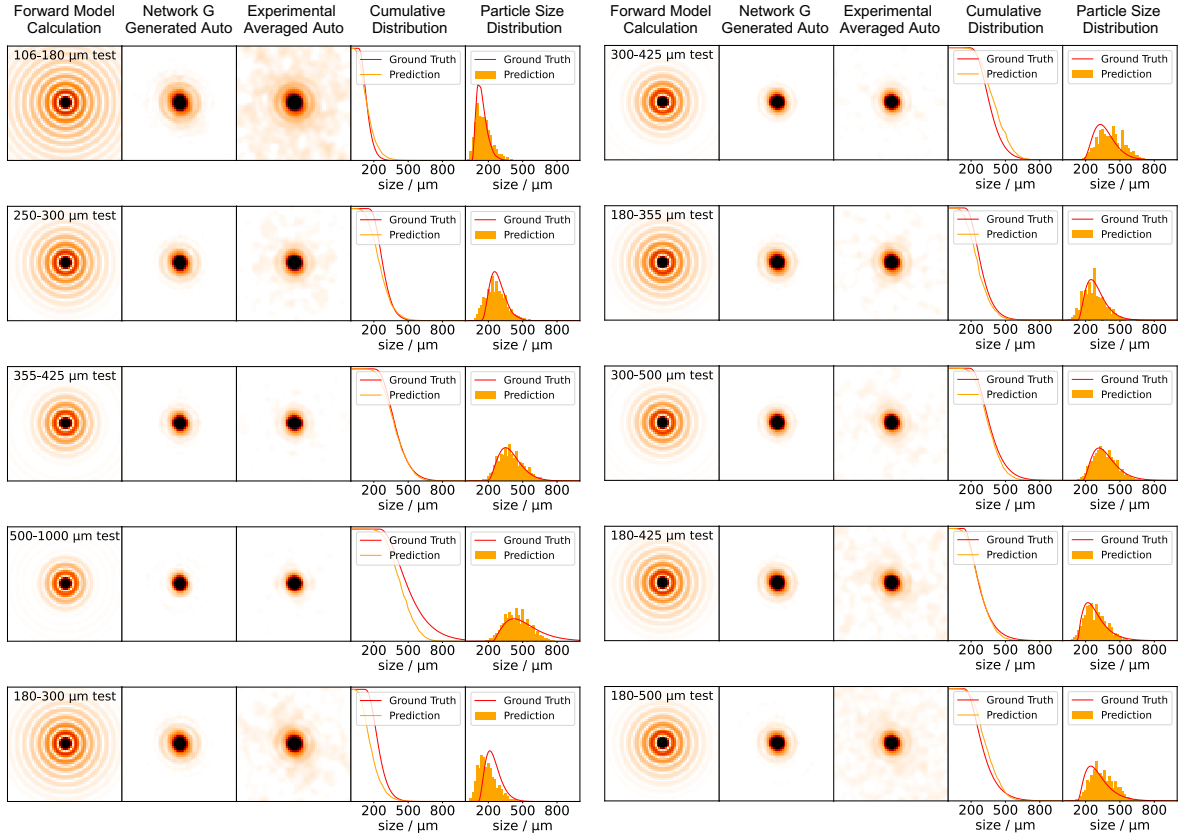

**Fig. 3. Prediction results with the experimental test dataset.** The results of 10 test samples are shown. The first and second columns show the output from the forward operator H and the generator G, respectively. The measured 200 frames averaged autocorrelations are plotted in the third column. The prediction results from the estimator F and ground truth (marked as red) of the cumulative distribution and the corresponding PSDs are plotted together in columns 4 and 5.

## Test Results and Model Visualization for PEACE

We conducted a comprehensive analysis of the generalization ability of the algorithm, shown in Fig. 3. This experimental test dataset was disjoint from the training data. The first column is the calculation result for the proposed forward model. The second column shows the image produced by the generator, which should match the corresponding ground truth, *i.e.* the experimental averaged autocorrelation. These images are averaged from 200 autocorrelation images, different from the 1000 frames averaged one in Fig. 2. Reducing the number of averaging frames sacrifices the signal-to-noise ratio but speeds up the data acquisition and prediction. There is still a slight mismatch between the generated image and the measured image in the surrounding area. These noise shape features are the residual from the ensemble average because the image is averaged with a finite number of frames only. Since they are away from the region of interest (later discussed in Fig. 4e), this deviation will not affect the performance of the estimator. The predictions and the ground truths measured from the particle size analyzer are plotted together in columns 4 and 5. The cumulative distribution follows from a clear physical definition of the size distribution for the non-spherical particles, such as the cubic-shaped KCl particles shown in Fig. 2a. More details about the cumulative distribution are included in Supplementary Section 4.

We use two methods to visualize our estimator. The first is plotting the output of each stage (the estimator's detailed structure is in Supplementary Section 4). As shown in Fig. 4a-d, the first stage output deemphasizes the zero-order peak and differentiates the side region. The second stage enhances the differentiation in the side area. The last two stages decode the information and extract the features.

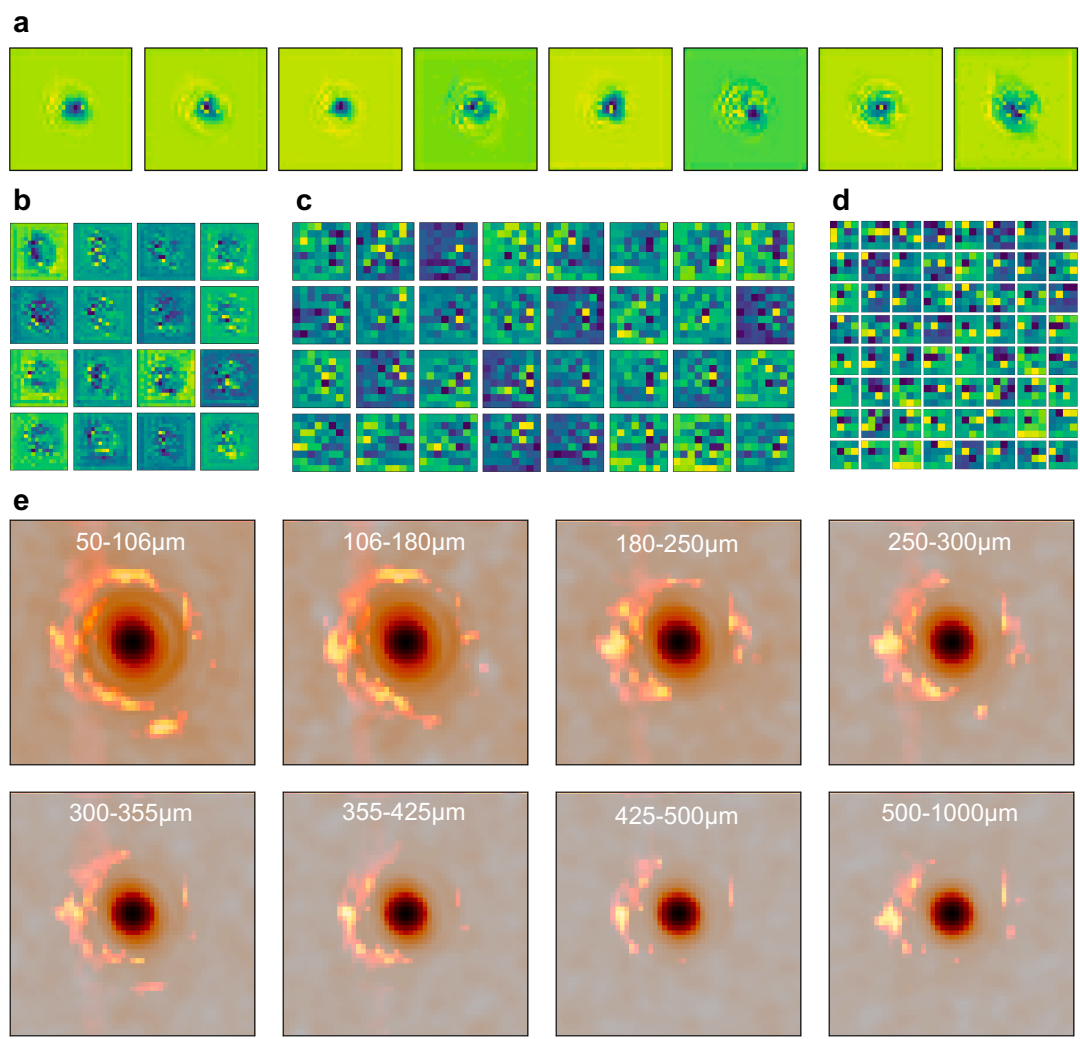

**Fig. 4 | Network visualization of the estimator F. a** The output of the first stage with 8 filters. **b** The output of the second stage with 16 filters. **c** The output of the third stage with 32 filters. **d** The output of the fourth stage with 64 filters. **e** GradCAM results for different sample sets**.** The input images are plotted together with the "importance" maps which consist of the flame shape features.

The other method is the Gradient-weighted Class Activation Mapping (GradCAM)[43]. This method visualizes the regions of input that are "important" for predictions. The results for different samples are plotted in Fig. 4e. For small-size samples, the attention is spread over the high-order lobes. For the big-size samples, the attention is concentrated on the second-order lobes since the higher-order lobes are no longer distinguishable. The attention for all plots avoids the central peak region, which is consistent with our theory.

**Time-lapsed PSD Monitoring in the Drying Process**

To validate the real-time monitoring applicability of our method, we carried out a time-lapse PSD measurement of an entire filter drying process[44]. The detailed information of our dryer is included in Supplementary Section 1. This demo process operated on 280g of KCl powder with a mixed solvent (water 40g/Ethanol 60g). Throughout the process, we maintained conditions of temperature 26 °C, pressure −720 mbar and agitation speed 4 rpm. Fig. 5b shows the PSD map *vs.* time. The sampling period of the PSD measurement was 15 seconds, including the data collection time for 200 frames and the computation time. The computer used for this measurement was an Intel Xeon W2245 CPU, 64 GB RAM, and NVIDIA Quadro RTX 5000 GPU with 64 GB VRAM.

For the duration of 5 to 25 minutes since the beginning of the process, we observed a gradual size increase compared to the original PSD. Since no crystallization or crystal growth could take place during drying, we think soft agglomeration instead occurred. From 25 to 40 minutes, the PSD gradually decreased to the original distribution, which is indicative of deagglomeration. At the far end (>75 mins), crystal breakage[45] caused the size distribution to decrease slightly compared to the original. PSD curves sampled at different times are plotted in Fig. 5a, together with Mastersizer results serving as ground truth for the beginning and ending time. They match well and the slight crystal breakage is clearly resolved for both Mastersizer and speckle predictions. The PSD curve shifts toward the right from the start time to 26.1 minutes, and then decreases backwards to the original position as plotted for the next 59.2 min. We did not measure the PSD with the Mastersizer when the powder was wet because in this process soft agglomerates were too fragile to allow any mechanical contact. This is one of the advantages of our non-invasive measurement compared to the traditional way with the Mastersizer. Fig. 5c are camera photos, taken with an iPhone 11, corresponding to the marked times in (a) and (b) for reference. (i) and (iv) are dry powders, while the (ii) and (iii) are wet powder images. It is hard to distinguish different sizes from macroscopic textures.

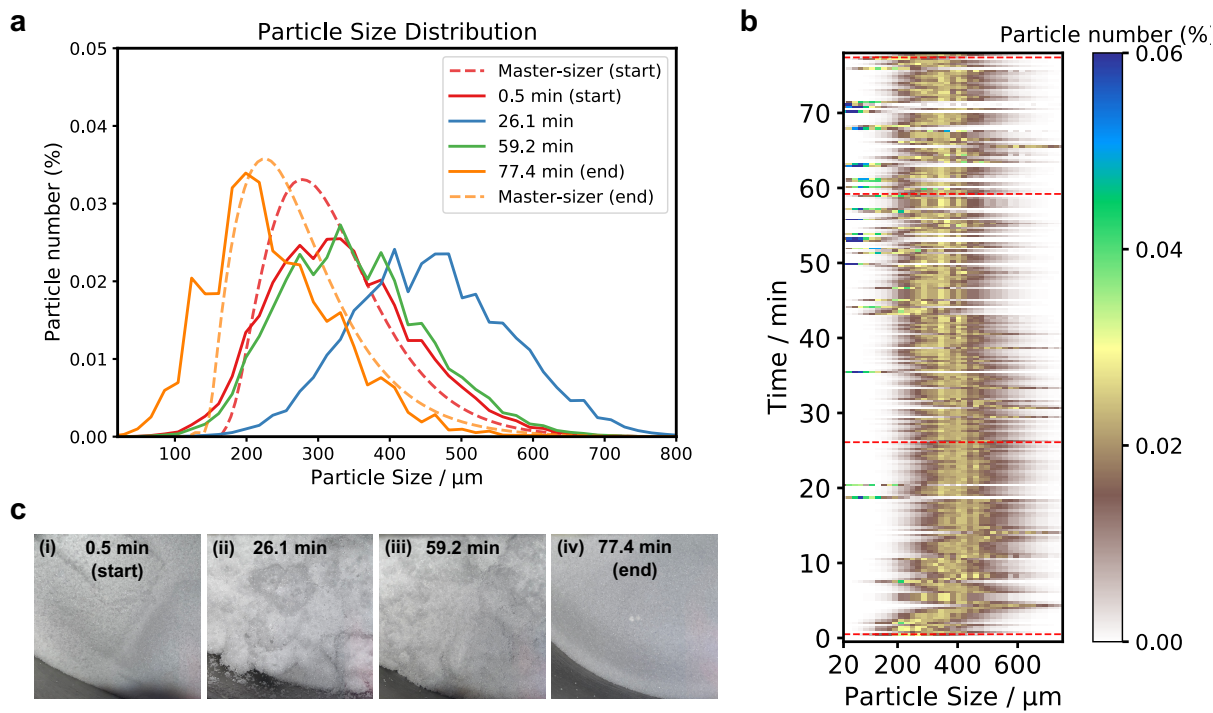

**Fig. 5 | Time-lapse PSD measurement in the drying process. a** Measured PSDs at a few selected time points corresponding to the dashed red line marks in **b**. The dashed line plots are the Mastersizer measurements serving as ground truth at the beginning and the end of the process. **b** Time-lapse PSD map during the drying process. We observed soft agglomeration occurring from 5-25 min, and deagglomeration from 25-40 min. **c** Corresponding camera photos at the times where the PSDs were sampled in **a**. These photos were taken from the same optical window as the laser beam using a commercial iPhone 11.

## Discussion

We have decoded the size information in the speckle pattern quantitatively with the help of the physics-enhanced autocorrelation-based estimator (PEACE). The data flow goes through a sequence of two neural networks, the generator and estimator, both informed by the law that we discovered relating the PSD to autocorrelation side lobes. Physics is involved in this method in four aspects. (1) The theoretical model guides the preprocessing of the raw speckle image, which is the ensemble-averaged autocorrelation (Supplementary Section 3). (2) The forward operator joins the PEACE training loop explicitly. (3) The generator's structure is inspired by the physics picture, resulting in high performance with only a few parameters, which is crucial for avoiding overfitting (Supplementary Section 4). (4) The output of the estimator is the cumulative size distribution, which is applicable to different particle shapes. Test results on needle shape materials, along with a theoretical discussion of the relationship between non-spherical particles and cumulative distributions are included in Supplementary Section 4.

One advantage of this physics-informed strategy is interpretability. The activation map results match our forward model very well, which means the size information is stored in the side lobes, and that enables an evaluation of the performance of the neural network in addition to the loss metric. For example, we want the generator's output to better match the ground truth at the 2nd ~ 5th lobes region. The center and the surrounding area are not very important. It is also possible to estimate the bound of the prediction ability from the forward model. Based on our theory, we even know how to push the prediction range into other regions of interest by tuning the parameters of the optical system. This is further discussed in Supplementary Section 5.

Our estimator fails to predict the double-peak shape PSD in a stress test with the synthetic data, yet it is able to estimate the appearance of large particles from the continuous monitoring –which is the most important for process control. This observation can be explained with the forward model; the results and explanation of the stress test are in Supplementary Section 6. This PEACE method is well suited for the extension to other contexts, including processes in the pharmaceutical industry such as blending and milling.

## Methods

**Optics Apparatus.** Our far-field optical probe can easily combine with an agitated filter dryer (AFD) typical of those used in the pharmaceutical industry. The wet solid is sealed in the dryer and the laser beam is delivered through a glass window. The [Fig. S2a](#) is a picture of the optics apparatus as we implemented it. For safety, the entire beam path except the output port facing the window is enclosed in an optical cage and optical tubes to prevent scattered light from escaping. This structure is compact, portable and, therefore, easy to transfer among different dryer systems without realignment. An agitator placed along the axis of the container impels the sample at a rotation speed of 4 rpm. The laser operates at 532 nm wavelength, and it is fairly easy to replace the source with a different wavelength, if desired. The angle of incidence on the surface is chosen to be approximately 10 degrees, so as to avoid specular back-reflection from the window onto the camera.

The optical beam path is shown in [Fig. S2b.](#) The laser beam is expanded to 4.8 mm with a beam expander in the telescope configuration, consisting of lenses L1 (focal length 25 mm) and L2 (30 mm). The beam is initially polarized in the direction parallel to the plane shown in the diagram, and so it is reflected by the polarizing beam splitter (PBS) to reach the sample inside the dryer through the glass window. Passage twice through the quarter-wave plate rotates the outgoing polarization direction from parallel to perpendicular so that the scattered light from the potassium chloride (KCl) powder sample is now transmitted through the PBS and propagates vertically upwards. The wave plate is also tilted by approximately 10 degrees for the same reason as the beam, to minimize specular back-reflection. Lens L3 (250 mm) concentrates the scattered light so that the CCD (model ZWO ASI183MM Pro) can capture an angular range as extensively as possible. More information about our apparatus can be found in Supplementary Section 1.

**Forward Model Theory.** The theory of the forward problem is established to link from a particular particle distribution to the raw speckle. The formulation of the forward problem is necessarily stochastic, treating the PSD as a probability density function which, in turn, determines the ensemble autocorrelation function of the raw speckle. The sketch for the initial simple analytical model is shown in Fig. S2c. We assume that only particles can scatter the light. Without loss of generality, the reflectivity is set to 1. In other words, the sketch considers the particles as a thin optical mask, which we denote as $a(x)$.

We interpret the particle radius $r_i$ as a random variable distributed according to the PSD $p(r)$. The particle location $x_i$ is also a random variable uniformly distributed across the object plane. $H(x)$ describes the surface height resulting from the randomly placed particles. The corresponding phase of the scattered light is $w(x) = \exp\left(j\frac{2\pi}{\lambda}H(x)\right)$. In the Fourier plane of L3, the electric field $E(x)$ is

$$E(x) = \int e^{j\frac{2\pi}{\lambda f_3}x\xi} S(\xi) d\xi, \qquad (2)$$

$$\text{where} \quad S(\xi) = a(x)w(x), \qquad (3)$$

and $f_3$ is the focal length of L3. The intensity collected by the CCD camera is

$$I(x) = |E(x)|^2 = \iint e^{j\frac{2\pi}{\lambda f_3}x(\xi_1 - \xi_2)} S(\xi_1) S^*(\xi_2) d\xi_1 d\xi_2. \qquad (4)$$

We now define the spatial-integral autocorrelation of the speckle image as

$$A(u') = \int I(x)I(x+u')dx. \qquad (5)$$

This equation may be rewritten in the form

$$A(u) = \left|\sum_i \frac{\sin(r_i u)}{u} e^{j2\pi x_i u}\right|^2. \qquad (6)$$

Here, $u = \frac{u'}{\lambda f_3}$, $i$ is the index of the $i$th particle, and $r_i$ and $x_i$ are the radius and the position for the $i$th particle, respectively. The full derivation leading from (5) to (6) is in Supplementary Section 3, equations

(M1)–(M14). If a sufficient number of particles find themselves within the field of view, then equation (6) can be reformulated as

$$A(u) = \left| \int dr \, p(r) \frac{\sin(ru)}{u} \sum_i e^{j2\pi x_i u} \right|^2. \quad (7)$$

This expression corresponds to the images in Fig. 2d. The granular feature results from the term $\sum_i e^{j2\pi x_i u}$ rather than measurement noise. The position information is encoded in this summation term. Since the particle coordinate $x$ is immaterial, we may eliminate it by ensemble-averaging the autocorrelation $A(u)$. Starting from equation (7), $\langle A(u) \rangle$ becomes

$$\langle A(u) \rangle = \frac{4 \sin^2\left(\frac{Du}{2}\right)}{D^2 u^2} \left| \int p(r) \frac{\sin(ru)}{u} dr \right|^2. \quad (8)$$

The details are in the Supplementary Section 3. In (8), $\langle \cdot \rangle$ denotes the ensemble average and $D$ is the beam spot diameter. This average operation cannot suppress the granular feature directly, but our physics model reveals that the $\sum_i e^{j2\pi x_i u}$ term is averaged into the $\frac{4 \sin^2\left(\frac{Du}{2}\right)}{D^2 u^2}$ term, which is irrelevant to the specific position distribution and only depends on the beam spot size. In this way, we change the stochastic expression (6) into a deterministic expression (8) for a given PSD, which ensures that the supervised learning is able to capture the correct map from the calculation result to the experimental result. Equation (8) is an intuitive yet approximate forward model between the raw speckle images and the PSD $p(r)$ through the averaged speckle autocorrelation function $\langle A(u) \rangle$.

## Data Availability

All processed data in this study have been deposited in the Harvard Dataverse under accession https://dataverse.harvard.edu/dataset.xhtml?persistentId=doi:10.7910/DVN/FZUG9V. The raw data is too big to upload into the public repository, please contact the author to request it. Line plots in Fig.2 and Fig.5 are provided in the Source Data file.

## Code Availability

All original codes have been deposited at https://github.com/qhzhang95/PEACE_Speckle and is publicly available (DOI: 10.5281/zenodo.7497506).

**Acknowledgements**

The authors acknowledge the MIT SuperCloud for providing HPC resources that have contributed to the research results reported within this paper/report. This research was supported by Millennium Pharmaceuticals, Inc. (a subsidiary of Takeda Pharmaceuticals), grant No. D824/MT15 (A.M., G.B. and R.B.).

**Author contributions**

G.B. and Q.Z. conceived the project and developed the theory. Q.Z. and J.G. designed the optical experiment and collected the data. Q.Z. analyzed the speckle data and developed the algorithm. J.G., A.P., Q.Z., W.T., C.P., C.M., and Y.Y. designed the filter dryer set up. Q.Z. and A.P. carried out the time-lapsed PSD drying experiment. Q.Z., J.G., and G.B. prepared the original manuscript and W.T. helped in the manuscript writing. C.P., C.M., Y.Y., M.S., and T.O. also contributed to the manuscript. G.B., A.M., and R.B. supervised this project.

**Competing interests**

The authors declare the following competing interests.

Patent

Patent applicant: Massachusetts Institute of Technology. Name of inventor(s): George Barbastathis, Qihang Zhang, Janaka C Gamekkanda Gamaethige, Richard D Braatz, Allan S Myerson. Application number: PCT/US22/50045. Status of application: Filed. Specific aspect of manuscript covered in patent application: System and method for determination of particle size distributions in this manuscript are all covered in the patent.

# From Laser Speckle to Particle Size Distribution in drying powders: A Physics Enhanced AutoCorrelation-based Estimator (PEACE)

# Supplementary Material

**Author list**

Qihang Zhang[1], Janaka C Gamekkanda[2], Ajinkya Pandit[2], Wenlong Tang[3], Charles Papageorgiou[4], Chris Mitchell[4], Yihui Yang[4], Michael Schwaerzler[5], Tolutola Oyetunde[5], Richard D Braatz[2], Allan S Myerson[2], and George Barbastathis[6,7]*

1. Department of Electrical Engineering and Computer Science, Massachusetts Institute of Technology, Cambridge, Massachusetts 02139, USA.

2. Department of Chemical Engineering, Massachusetts Institute of Technology, Cambridge, Massachusetts 02139, USA.

3.Data Sciences Institutes, Takeda Pharmaceuticals International Co, 650 E Kendall St, Cambridge, Massachusetts 02142, USA.

4. Process Chemistry Development, Takeda Pharmaceuticals International Co, 40 Landsdowne St, Cambridge, Massachusetts 02139, USA.

5. Innovation and Technology Sciences, Takeda Pharmaceutical Company Limited, 200 Shire Way, Lexington, MA 02421

6. Department of Mechanical Engineering, Massachusetts Institute of Technology, Cambridge, Massachusetts 02139, USA.

7. Singapore-MIT Alliance for Research and Technology (SMART) Centre, 1 Create Way, Singapore 117543, Singapore.

* Email: gbarb@mit.edu

## 1. Apparatus overview.

**Filter dryer overview.** The filter drying device shown in Fig. S1 was designed based on a prototype[1]. The device has 1400 ml capacity with overall device dimensions of 330 mm height × 220 mm width × 220 mm length without the optical components. The body, impeller, and filter mesh are made of steel, and the lid is made of aluminum. The lid contains a vacuum port, an air/gas input port, a camera port, two wash solvent ports, a feed suspension port, and a glass observation window. A vacuum gauge and pressure release valve are attached to the lid. The impeller is connected to a motor on the top of the lid. The impeller blade is designed with teeth to promote cake depumping. An inductive heater and a thermocouple are attached to the bottom of the device to control the drying temperature. This inductive heating plate is used to heat the steel base and the steel body of the dryer. An endoscope camera is attached to the lid to record process videos of the drying material from the top. The optics bench sheds light on the sample and collects the scattered beam through the observation window.

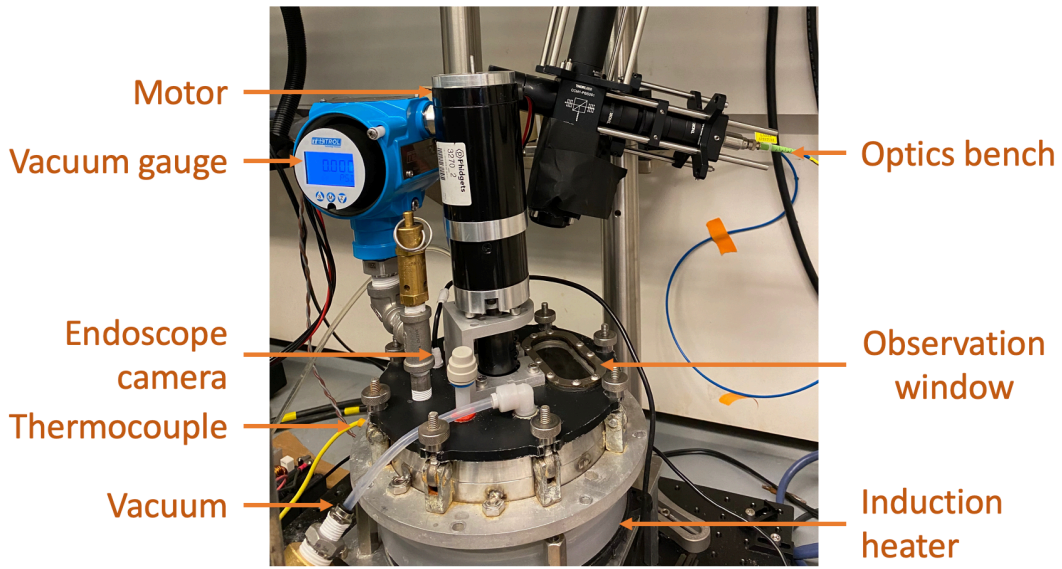

**Fig. S1 | The photo of our drying filter.** The crucial parts are labeled.

**Optics bench.** The laser model is Excelsior 532 Single Mode with 300 mW output power. After the fiber coupling, only 230 mW power injects into our system. Our lenses are made of N-BK7 with anti-reflection coating, whose transmission is 99.5% at 532 nm. The beam has a polarization ratio higher than 100:1. The extinction ratio of the polarized beam splitter (PBS) is higher than 1000:1. The transmission of the quarter plate and the silica window are 99.8% and 90%, respectively. The beam power on the sample surface is 203 mW. The filter dryer contains 150g KCl powder with a 690 J/kg·K specific heat capacity. From this calculation, our laser induces a 2.4 mK/s temperature rising rate, which is low enough not to disturb the drying process for the duration of observation. The power of the scattered light measured at the CCD plane is 120 µW. The coherence length of the laser is 25 m, corresponding to a temporal bandwidth of 0.01 pm. It provide sufficient temporal coherence to produce sharp speckles.

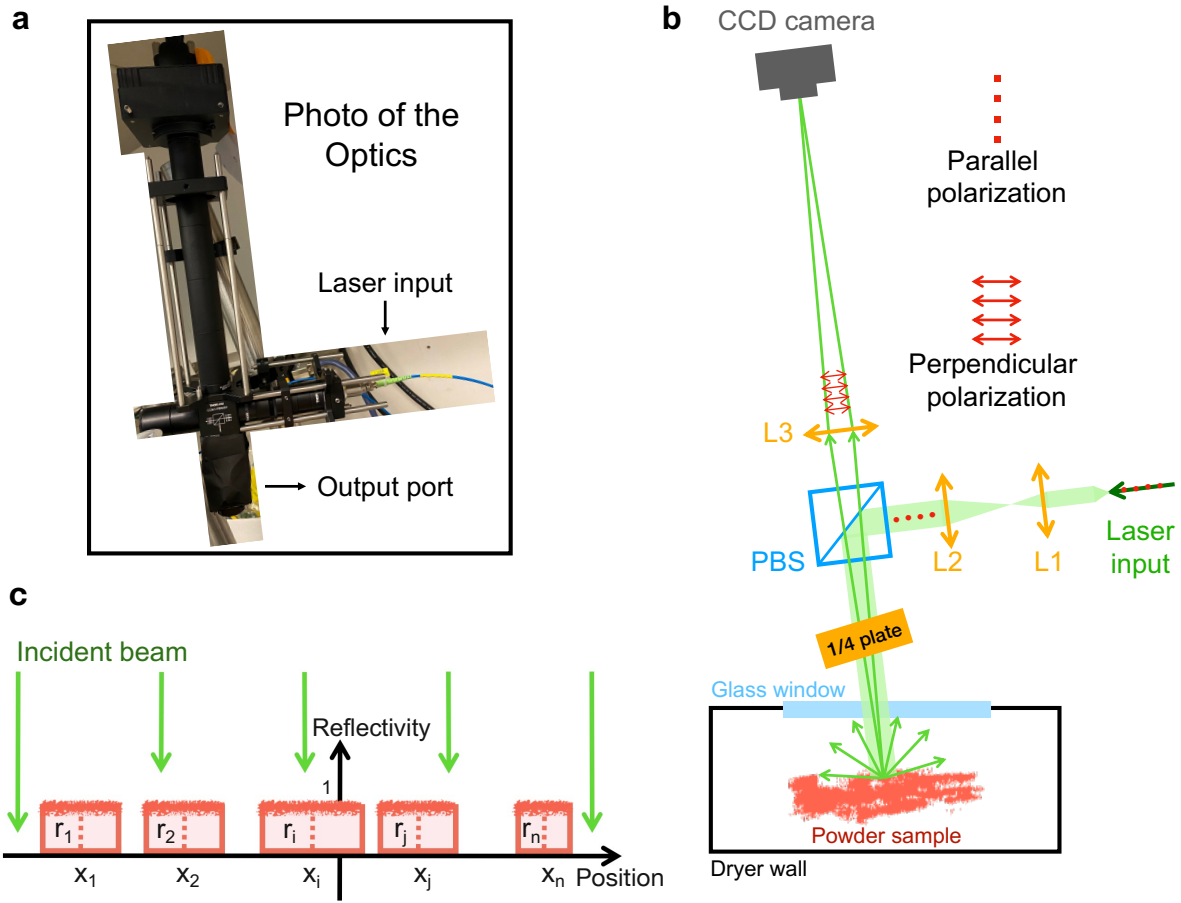

**Fig. S2** | **(a)** A photo of our optics bench in the lab. All optical components are encapsulated in the optical tube and cubic to prevent light from escaping through the side. **(b)** A sketch of the optics. Lenses L1, L2, and L3 have focal lengths of 25 mm, 30 mm, and 250 mm, respectively. **(c)** A powder model sketch used in the derivation of the forward model. We only consider the top surface. The red rectangles are 1D powder particles, with radius $r_i$ and position $x_i$. The rough top edge denotes that the scattered beam has a random phase at each different position.

**Image acquisition.** Our monochromatic CCD model is ZWO ASI183MM Pro with 5496 × 3672 pixels with a 2.4 μm pixel size. We run it in the bin-pixel mode with 1920 × 1080 pixels and 70 fps framerate. The typical size of the speckle pattern calculated from $\lambda f/D$ is 28 μm for our imaging system, which is larger than the pixel size to ensure a good resolution of the speckle pattern. We crop the central uniformly illuminated area to 1024 × 1024 pixels. Fig. S3 shows the single-frame raw speckle images and the corresponding autocorrelations with different exposure times. We find that 100 μs and 200 μs exposure times maintain a high degree of spatial correlation, whereas the speckle begins to decorrelate at exposures in excess of 500 μs. At exposure as high as 1 ms, the speckle pattern blurs and the corresponding autocorrelation disperses. We choose 100μs as our exposure time to ensure that it is short enough to maintain the speckle spatial correlation within each frame. The time scale of PSD evolution in the powder drying dynamics is tens of minutes (as shown in Fig. 5) which is much longer than the time of a single PSD measurement. So we can safely assume that the powder PSD does not change during the data collection time.

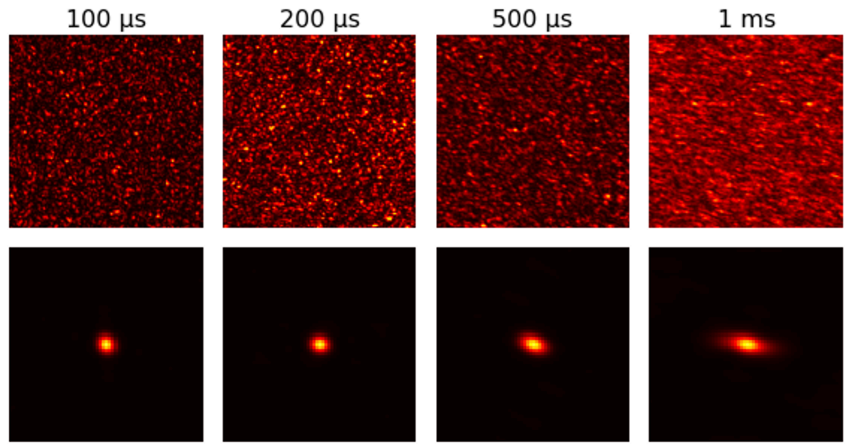

**Fig. S3 | Speckle images (top) and its autocorrelations (bottom) in different exposure times.**

## 2. Sample preparation and calibration.

Commercially obtained potassium chloride (KCl) powder was used to calibrate the speckle. The morphology of the KCl is blocked-shaped crystals. Twenty samples of KCl having different size distributions were prepared by sieving bulk KCl using sieves attached to the sieve shaker. The sieves were stacked in the sieve shaker in the order of decreasing sieve opening size from top to bottom (Sieve opening sizes used: 500 μm, 425 μm, 355 μm, 300 μm, 250 μm, 180 μm, 106 μm). The sieving was carried out 15-30 minutes until no further change of weight of the sieves with powder was observed. The samples obtained from the sieves were directly used to calibrate the speckle. Each sieved dry powder sample was added individually to the filter-dryer and stirring was carried out at 4 rpm to record 1000 frames speckle pattern. 20 calibration data sets were obtained corresponding to the sieved KCl samples.

From 1000 collected images in each sample set, we use 200 frames on average with a 40-frame step sliding window to obtain 20 averaged images. Employing an ergodicity argument, these averaged images are the ensemble averaged autocorrelations. To evaluate the generalization ability of our DNN model, we separate 10 sample sets to form the generator G training dataset, whereas the remaining 10 sets are the test dataset which is disjoint from the training process. Both datasets are kept away from the estimator F training process.

Offline particle size analysis was carried out by laser diffraction to obtain the ground truth particle size distribution of each KCl sample set. Malvern Mastersizer 2000 attached to a Scirocco 2000 dry dispersion unit was used to obtain particle size data. Fig. S4a shows the PSD data obtained from Mastersizer corresponding to each sieved sample. The finite sieves and master-sizer measurement limit the number of datasets we can collect, as they are time-consuming and unrecyclable.

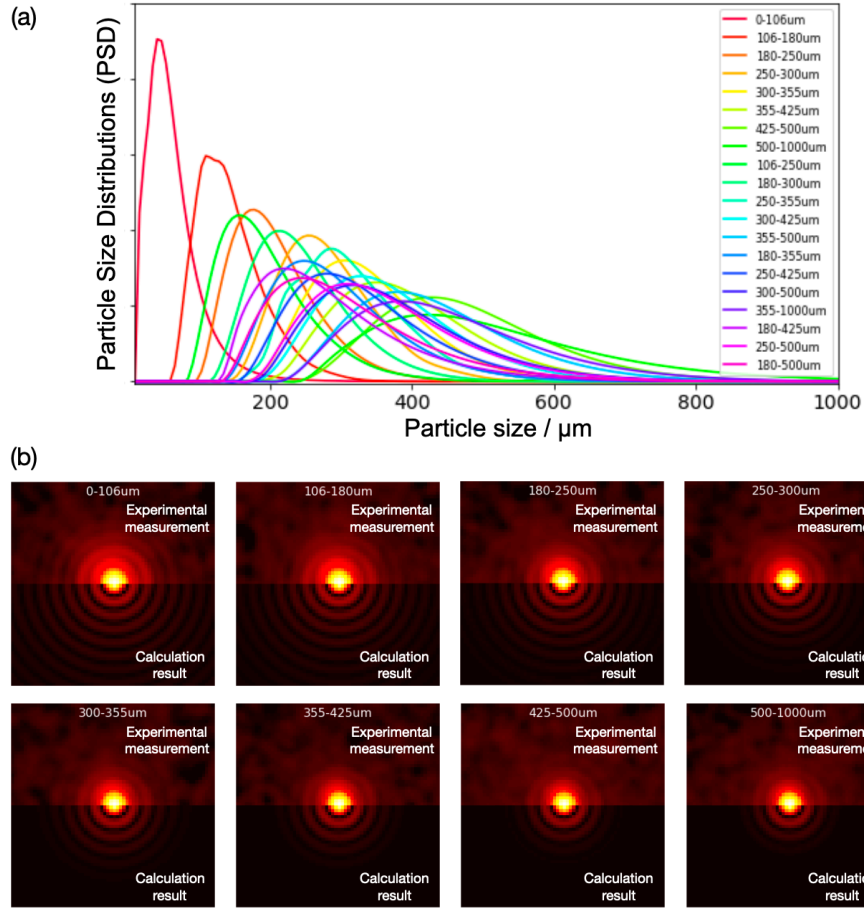

**Fig. S4 | Sample preparation and calibration. (a)** PSD calibration curves measured from the Mastersizer serve as the ground truth for each data set. **(b)** Measured average-autocorrelations and the calculations for 8 sample sets with ascending size distributions are plotted together. Their corresponding line-cut plots are shown in Fig.2(g).

## 3. Forward model.

The detailed math derivation of our forward model is described in this section. It continues from equation (4) in the Methods section. Substituting equations (2) and (3) into equation (4),

$$A(u') = \int \iint \iint e^{j\frac{2\pi}{\lambda f_3}[x(\xi_1-\xi_2)+(x+u')(\eta_1-\eta_2)]} S(\xi_1)S^*(\xi_2)S(\eta_1)S^*(\eta_2)dxd\xi_1d\xi_2d\eta_1d\eta_2 \quad (M1)$$

After integrating over $x$,

$$A(u') = \iint \iint e^{j\frac{2\pi}{\lambda f_3}u'(\eta_1-\eta_2)} S(\xi_1)S^*(\xi_2)S(\eta_1)S^*(\eta_2)d\xi_1d\xi_2d\eta_1d\eta_2 \int e^{j\frac{2\pi}{\lambda f_3}(\xi_1-\xi_2+\eta_1-\eta_2)x}dx$$

$$= \iint \iint e^{j\frac{2\pi}{\lambda f_3}u'(\eta_1-\eta_2)} S(\xi_1)S^*(\xi_2)S(\eta_1)S^*(\eta_2)\,\delta(\xi_1-\xi_2+\eta_1-\eta_2)d\xi_1d\xi_2d\eta_1d\eta_2 \quad (M2)$$

The delta function sifts the content of the integrand at $\xi_2 = \xi_1 + \eta_1 - \eta_2$. Then the equation can be further simplified as,

$$A(u') = \iint \iint e^{j\frac{2\pi}{\lambda f_3}u'(\eta_1-\eta_2)} S(\xi_1)S^*(\xi_1+\eta_1-\eta_2)S(\eta_1)S^*(\eta_2)d\xi_1d\eta_1d\eta_2 \quad (M3)$$

From Equation (M3), we carry out variable substitution as $\eta = \eta_2$, $\sigma = \eta_1 - \eta_2$ and $\tau = \xi_1 - \eta_2$.

$$A(u') = \iint \iint e^{j\frac{2\pi}{\lambda f_3}u'(\sigma+\eta-\eta)} S(\eta+\tau)S^*(\sigma+\eta+\tau)S(\eta+\sigma)S^*(\eta)d\tau d\sigma d\eta$$

$$= \int d\tau \left[\int e^{j\frac{2\pi}{\lambda f_3}u'(\sigma+\eta)} S^*(\sigma+\eta+\tau)S(\sigma+\eta)d\sigma\right]\left[\int e^{-j\frac{2\pi}{\lambda f_3}u'\eta} S(\eta+\tau)S^*(\eta)d\eta\right]$$

$$= \int d\tau \left[\int e^{j\frac{2\pi}{\lambda f_3}u'\sigma} S^*(\sigma+\tau)S(\sigma)d\sigma\right]\left[\int e^{-j\frac{2\pi}{\lambda f_3}u'\eta} S(\eta+\tau)S^*(\eta)d\eta\right]$$

$$= \int d\tau \left|\int e^{j\frac{2\pi}{\lambda f_3}u'\sigma} S(\sigma)S^*(\sigma+\tau)d\sigma\right|^2$$

$$(M4)$$

Let $u = \frac{2\pi u'}{\lambda f_3}$,

$$A(u) = \int d\tau \left|\int e^{ju\sigma} S(\sigma)S^*(\sigma+\tau)d\sigma\right|^2 \quad (M5)$$

After substituting $S$ from equation (2),

$$A(u) = \int d\tau \left|\int e^{ju\sigma} a(\sigma)a^*(\sigma+\tau)w(\sigma)w^*(\sigma+\tau)d\sigma\right|^2 \quad (M6)$$

In the internal integral over $\sigma$, the phase term $w(\sigma) = \exp\left(j\frac{2\pi}{\lambda}H(\sigma)\right)$ varies much faster than $a(\sigma)$ nd $e^{ju\sigma}$, so we can apply the rotating wave approximation to move $w(\sigma)w^*(\sigma+\tau)$ outside the integral over $\sigma$.

$$A(u) = \int d\tau\, W(\tau) \left| \int e^{ju\sigma} a(\sigma)a^*(\sigma+\tau)d\sigma \right|^2, \tag{M7}$$

where $W(\tau) = \langle w(\sigma)w^*(\sigma+\tau)\rangle_\sigma$ is the spatial average of $w(\sigma)w^*(\sigma+\tau)$ over $\sigma$. $W(\tau)$ describes spatial correlation of the surface phase. $W(\tau)$ will drop to 0 if $\tau$ is larger than the correlation length. For the ideal rough surface, the correlation length is infinitely small and $W(\tau)$ degrades into $\delta(\tau)$.

$a(x)$ is the mask defined by the particles, and we may express it as,

$$a(\sigma) = \sum_i \text{Rect}\left(\frac{\sigma - x_i}{r_i}\right) \tag{M8}$$

$$a(\sigma)a^*(\sigma+\tau) = \sum_{i,j} \text{Rect}\left(\frac{\sigma - x_i}{r_i}\right)\text{Rect}\left(\frac{\sigma - (x_j - \tau)}{r_j}\right) \tag{M9}$$

Where $\text{Rect}(x) = 1$ when $x \in [-1,1]$, otherwise $= 0$ is the boxcar function. Since the particles cannot overlap, and the correlation length is well smaller than the particle size, we can approximately assume $\text{Rect}\left(\frac{\sigma - x_i}{r_i}\right)\text{Rect}\left(\frac{\sigma - (x_j - \tau)}{r_j}\right) = 0$ for $i \neq j$.

$$a(\sigma)a^*(\sigma+\tau) = \sum_i \text{Rect}\left(\frac{\sigma - x_i}{r_i}\right)\text{Rect}\left(\frac{\sigma - (x_i - \tau)}{r_i}\right) = \sum_i \text{Rect}\left(\frac{\sigma - (x_i + \frac{\tau}{2})}{r_i - \frac{\tau}{2}}\right) \tag{M10}$$

After substituting equation (M10) to equation (M7), we obtain

$$A(u) = \int d\tau\, W(\tau) \left| \sum_i \int e^{ju\sigma} \text{Rect}\left(\frac{\sigma - (x_i + \frac{\tau}{2})}{r_i - \frac{\tau}{2}}\right) d\sigma \right|^2 \tag{M11}$$

The internal integral is the Fourier transform

$$\int e^{ju\sigma} \text{Rect}\left(\frac{\sigma - (x_i + \frac{\tau}{2})}{r_i - \frac{\tau}{2}}\right) d\sigma = e^{-ju\left(x_i+\frac{\tau}{2}\right)} \frac{\sin\left(u\left(r_i - \frac{\tau}{2}\right)\right)}{u}. \tag{M12}$$

Substituting, we find

$$A(u) = \int d\tau\, W(\tau) \left| \sum_i e^{-ju\left(x_i+\frac{\tau}{2}\right)} \frac{\sin\left(u\left(r_i - \frac{\tau}{2}\right)\right)}{u} \right|^2. \tag{M13}$$

If we apply the assumption that $W(\tau) = \delta(\tau)$, then equation (M13) may be transformed as

$$A(u,t) = \left| \sum_i \frac{\sin(r_i u)}{u} e^{j2\pi x_i(t)u} \right|^2. \tag{M14}$$

Here, $u = \frac{u'}{\lambda f_3}$, $i$ is the index of the $i$-th particle, and $r_i$ and $x_i$ are the radius and the position for the $i$-th particle, respectively. This is the same as equation (5) in the main text.

Below are the derivations of the ensemble average of the autocorrelation $A(u,t)$. The ensemble average is over independent measurements. Specifically, it is an average over $x_i$. However, $x_i$ is a function of time $t$ in our system since the powder are agitated by the impeller. Since $x_i(t)$ is ergodic, we may replace the average of measurements at different time $t$ with the ensemble average. Starting from equation (M14), and since $W(\tau)$ is invariant, we may move the ensemble average bracket into the integral, as

$$\langle A(u) \rangle_t = \int d\tau\, W(\tau) \left\langle \left| \sum_i e^{-ju\left(x_i(t)+\frac{\tau}{2}\right)} \frac{\sin\left(u\left(r_i - \frac{\tau}{2}\right)\right)}{u} \right|^2 \right\rangle_t$$

$$= \int d\tau\, W(\tau) \left\langle \sum_{i,j} e^{-ju\left[(x_i(t)+\frac{\tau}{2}) - (x_j(t)+\frac{\tau}{2})\right]} \frac{\sin\left(u\left(r_i - \frac{\tau}{2}\right)\right)}{u} \frac{\sin\left(u\left(r_j - \frac{\tau}{2}\right)\right)}{u} \right\rangle_t$$

$$= \int d\tau\, W(\tau) \int p(r_1) p(r_2) \frac{\sin\left(u\left(r_1 - \frac{\tau}{2}\right)\right)}{u} \frac{\sin\left(u\left(r_2 - \frac{\tau}{2}\right)\right)}{u} dr_1 dr_2 \left\langle \sum_{i,j} e^{-ju(x_i(t)-x_j(t))} \right\rangle_t$$

$$\tag{M15}$$

The third equal sign comes from the fact that $r$ and $x$ are independent. The radius $r_i$ follows the probability distribution $p(r)$, which is invariant. The $x_i$'s are randomly distributed in space, forming an ergodic process. Therefore, the ensemble average over $x$ is equal to the spatial average:

$$\left\langle \sum_{i,j} e^{-ju(x_i(t)-x_j(t))} \right\rangle_t = \frac{1}{D^2} \int_{-\frac{D}{2}}^{\frac{D}{2}} e^{-ju(x_1-x_2)} dx_1 dx_2 = \frac{4\sin^2\frac{Du}{2}}{D^2 u^2}, \tag{M16}$$

Where $D$ is the laser spot diameter. The term $\frac{Du}{2} = \pi$ term determines the average speckle size, which is consistent with the textbook[2] result.

$$\int p(r_1)p(r_2) \frac{\sin\left(u\left(r_1-\frac{\tau}{2}\right)\right)}{u} \frac{\sin\left(u\left(r_2-\frac{\tau}{2}\right)\right)}{u} dr_1 dr_2 = \left|\int p(r) \frac{\sin\left(u\left(r-\frac{\tau}{2}\right)\right)}{u} dr\right|^2 \quad (M17)$$

If we substitute equations (M16) and (M17) back to equation (M15), we obtain

$$\langle A(u)\rangle = \frac{4\sin^2\frac{Du}{2}}{D^2 u^2} \int d\tau\, W(\tau) \left|\int p(r) \frac{\sin\left(u\left(r-\frac{\tau}{2}\right)\right)}{u} dr\right|^2 \quad (M18)$$

The particle size of our sample varies from ~50μm to ~1000μm, which is much larger than the wavelength 532nm. Thus, we meet the criterion $\lambda \ll H(x)$, implying that the phase correlation length is much smaller than the particle size. This is validated from our simulation in Fig. S5, so we may adopt the assumption $W(\tau) = \delta(\tau)$ safely. Therefore, we obtain the final expression (the same as equation 1),

$$\langle A(u)\rangle = \frac{4\sin^2\left(\frac{Du}{2}\right)}{D^2 u^2} \left|\int p(r) \frac{\sin(ru)}{u} dr\right|^2. \quad (M19)$$

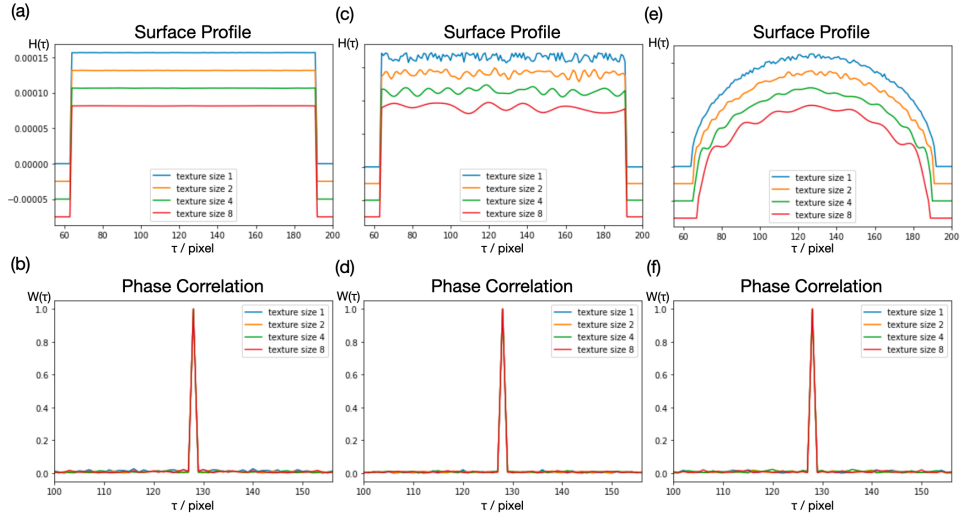

**Fig. S5 | Simulation of phase correlation for different surface height profiles. (a)** Surface profile for a square shape particle with a size ~150μm. The particle is sampled with 128 pixels. The height fluctuation is 1%, which corresponds to 1.5μm. Four curves with different texture sizes are plotted. The texture size describes the horizontal size of the fluctuations in pixel units. **(b)** plots the phase correlation calculated by $W(\tau) = \exp\left(j\frac{2\pi}{\lambda}H(\tau)\right)$ from the surface profile (a). **(c)** Surface profile with 10% height fluctuation. The rest information is the same with (a). **(d)** Phase correlation corresponding to (c). **(e)** Surface profile for a round particle with 5% height fluctuation. **(f)** The phase correlation corresponding to (e). The correlation lengths for all different height fluctuations and horizontal textures are smaller than a single pixel. Thus, we may safely assume that $W(\tau) \approx \delta(\tau)$.

# 4. Physics Enhanced AutoCorrelation-based Estimator (PEACE) algorithm

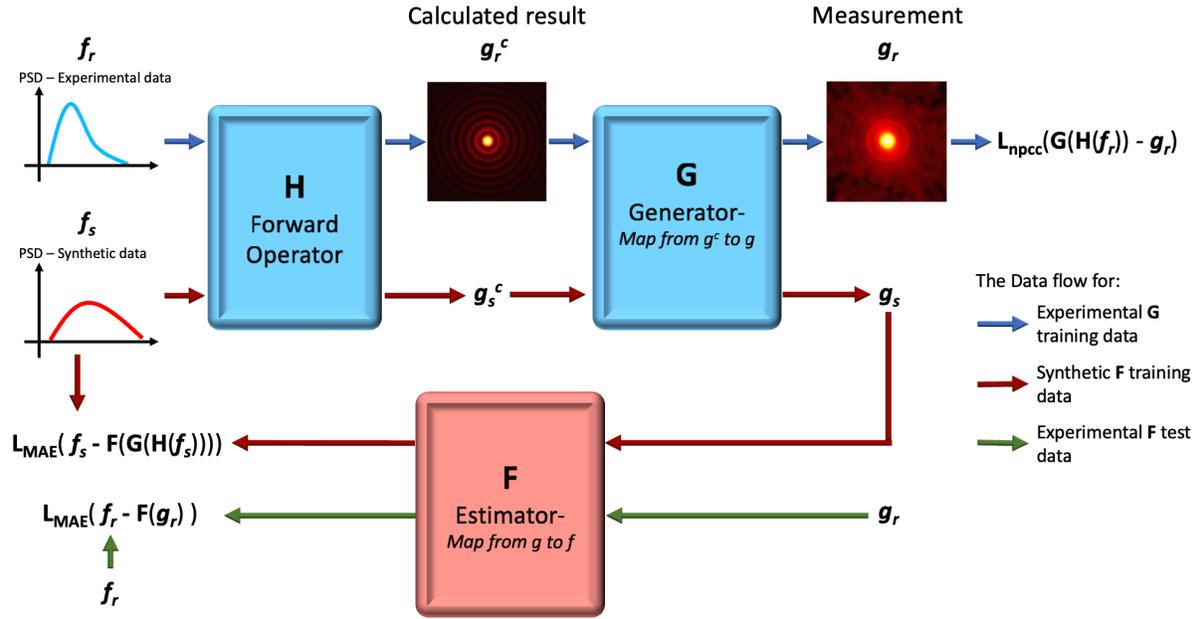

**Fig. S6 | The learning scheme of the PEACE algorithm.** A total of three data flows are described. First, an experimental dataset is utilized for training the generator with the NPCC loss function. Then a synthetic dataset is generated from the forward operator and the generator to train the estimator with the MAE loss. Finally, another experimental data serves as the test set for the estimator.

**PEACE learning scheme.** The forward relationship (1) is highly nonlinear; moreover, the analysis in Fig. 2 reveals that the inverse problem is also ill-conditioned: the relevant information for the PSD is in the sidelobes of the term $\frac{4\sin^2\left(\frac{uD}{2}\right)}{D^2 u^2}$ and, hence, the solution may be disproportionately susceptible to detection noise and other disturbances. As a general strategy to stabilize the solution against disturbances, we must employ some form of regularization. In typical situations, the most effective regularizing priors are derived from sparsity arguments (also referred to as compressed sensing[3]). For linear inverse problems, sparse formulations lead to convex functionals that are numerically dealt with using methods such as TwIST and ADMM[4,5].

Our inverse problem with equation (1) is somewhat atypical, and not only because it is nonlinear. The probability for a specific radius $p(r)$ is manifestly modulated by $\frac{\sin(ru)}{u}$. The term $\frac{\sin(ru)}{u}$ approaches 0 as r approaches 0, adding to the ill-posedness already imposed by the preceding term $\frac{4\sin^2\left(\frac{uD}{2}\right)}{D^2 u^2}$ term. This suggests that sparsifying the PSD as, for example, a superposition of radial basis functions, might be risky. Instead, we adopt a machine learning approach, using data to learn the regularizing prior. Since we can only provide finite sample sets to get data from our experiment and the independent particle size analyzer, it is not enough to train a large neural network acting in "estimator" capacity to solve the inverse problem independently. Instead, with the help of the forward model, it is sufficient to complement the forward operator by training a small neural network called a "generator." Then we can generate a large synthetic dataset from the forward process to meet the data requirement of the training process for the estimator. The final answer is to solve the forward and inverse problems collaboratively. This algorithm is named "the Physics Enhanced AutoCorrelation-based Estimator (PEACE)."

Fig. S6 shows the PEACE learning scheme. It consists of three parts, forward operator **H**, generator **G**, and estimator **F**. The forward operator **H** refers to equation 1, which produces the calculated result $g^c$ from a given object $f$, the PSD in our case. The generator **G** maps $g^c$ to the experimental measurement $g$. The estimator F is our ultimate final goal, which converts $g$ to the corresponding object $f$. A paired small training dataset ($g_r = \mathbf{H}(f_r)$ and $g_r^c$) is collected from the experiment to train the generator **G** with the negative Pearson correlation coefficient (NPCC)[6,7] loss function. The generator **G** is designed based on the physical model to reach high performance with only 2.8k parameters, thus avoiding overfitting. A large synthetic dataset ($g_s^c$ and $f_s$) is generated from **H** and **G** to train the estimator **F** with the mean-absolute-error (MAE) loss. A separated experimental dataset tests the performance of the estimator **F** on the real data.

**The structure of the generator G**. The generator contains 2.8k parameters. The generator compensates for the particles overlapping along the longitudinal direction and the deviation induced by the finite spatial integral and finite frames average in the ensemble autocorrelation calculation. It is designed based on the physical model. Random noise generated from a uniform distribution over [0,1) is concatenated to the input to mimic the fluctuation induced by the effect of finite averaging. Moreover, adding noise can facilitate the generalization to the measured style since in the experimental measurement noise is inevitably present. Our forward operator only considers the top layer particles. The generator handles the influence of the underlying layers. Light scattered from the second layer particles will partially fill the gap between the first layer particles, resulting in a larger "effective particle size". Since the second layer particles are also densely located in the plane, this "large effective size" should be comparable to the beam spot size (5mm), which is much larger than the typical particle size (100 μm-1000 μm), resulting in a substantial intensity reduction of the high-order lobes in the experiment compared to the calculation result only from the first layer. We assume that the multilayer effect is a blurring and non-linear function applied to the original calculation. This physical model can be written in the following expression,

$$g = \text{normalize}(\, g^c + h(g^c)\,) \qquad (M20)$$

where $h(\cdot)$ denotes the nonlinear and convolutional blur operation. With this inspiration, we design the structure of **G** in the following way,

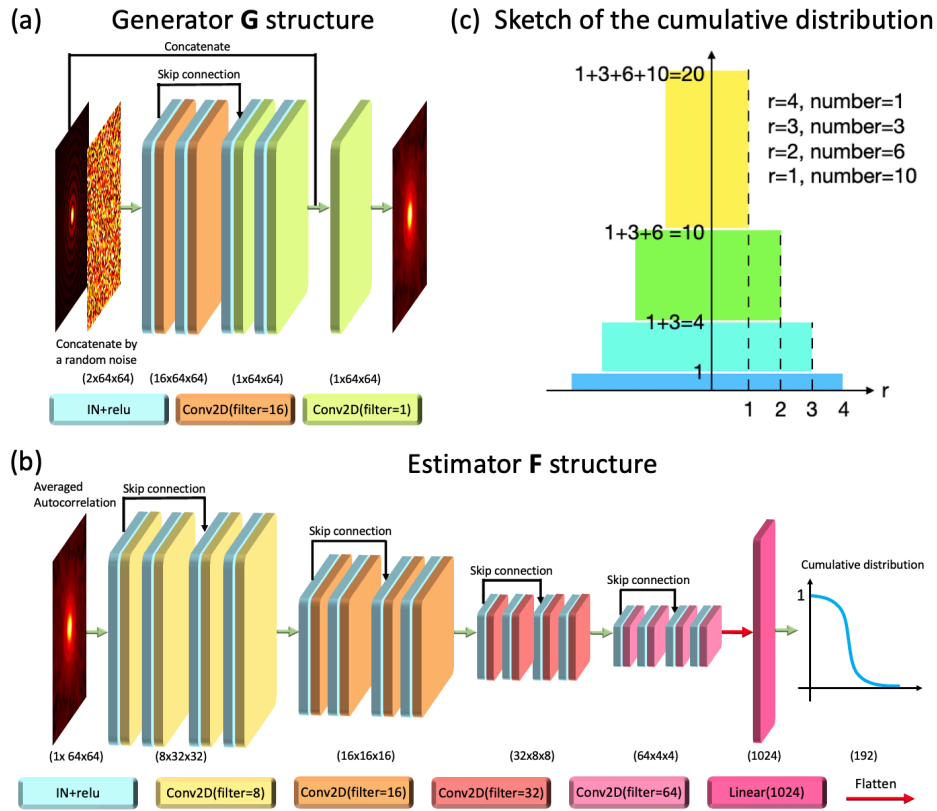

**Fig. S7 | (a) Generator structure.** The generator transforms the calculated image to the measured image. Random noise is generated and concatenated to the input. This neural network contains four convolutional layers. Instance-normalization (IN) and ReLU activation are applied before each convolutional layer except the last one. The last layer has a Sigmoid activation function. **(b) Estimator structure.** The estimator takes averaged autocorrelation images as input and the cumulative distributions as output. It consists of four stages with four convolutional layers in each. IN and ReLU activation are applied before each convolutional layer. There is a flattening layer and a linear layer with Sigmoid activation connected to the output of the fourth stage. The output data dimension and the filter parameters for each layer are labeled at the bottom of this figure. **(c)** This image is to show the physical meaning of the cumulative distribution. Assume there are 20 particles, ten particles have radius 1, six particles have radius 2, three particles have radius 3, and one particle has radius 4. If we stack them at the center and set the y-axis as the particle number, after normalization we find that the vertical "projections" are identical to the cumulative distribution.

$$g = \text{sigmoid}(g^c + \alpha \text{NN}(g^c) + \beta) \qquad (M21)$$

where the slope and bias constants $\alpha$, $\beta$ and the sigmoid function play the role of the last convolution layer with filter 1 and kernel size $1 \times 1$. The sigmoid function can scale the final output to $[0,1]$. The function $\text{NN}(\cdot)$ consists of four cascaded convolutional layers with $3 \times 3$ kernel and ReLU activation. It learns from data to fit $h(\cdot)$ without any downsampling. The detailed structure of **G** is shown in Fig. S7a. With this physics-inspired design, the generator **G** can reach high performance with only 2.8k parameters. This is a critical point to avoid overfitting with the small experimental training set.

**The structure of the estimator F**. The estimator contains 377k parameters. The structure of the estimator **F** is shown in detail in Fig. S7b. The ensemble averaged autocorrelation serves as the input. The estimator consists of four convolutional stages followed by a linear layer. Each stage has four 2D convolutional layers of kernel size 3, combined with Instance-Normalization layer (IN)[8] and ReLU activation. At each stage, the first convolution layer downsamples the image by a factor of 2. The filter depth increases gradually in each stage from 16 to 64. The skip connection structure serves to reduce the gradient vanishing effect[9]. After three convolution stages, the 3D signal is flattened and passed through a linear layer to the output, a non-parametric curve described by 192 samples. This is the cumulative distribution of the particle sizes, which we then differentiate to obtain the PSD. To reduce fluctuations, we bin the pixels of the PSD curve from 192 to 64 pixels.

There are two reasons why we pick up the cumulative distribution rather than the PSD directly as the output of the neural network. First, the cumulative distribution is monotonic from 0 to 1, which discourages overfitting the fluctuations that would inevitably appear in the PSD. Secondly, the cumulative distribution may easily be derived from Equation (1) as

$$\text{FT}\left(\int p(r) \frac{\sin(ru)}{u} dr\right)(x) = \int p(r) \text{Rect}\left(\frac{x}{r}\right) dr = \int_0^{|x|} p(r) dr \qquad (M22)$$

where $\text{Rect}(x) = 1$ when $x \in [-1,1]$, otherwise $= 0$ is the boxcar function. The physical meaning of the right-hand term in (M22) is shown in Fig. S7c. In effect, it moves all particles to the center, and the vertical axis becomes the number of particles as a function of $r$. After normalization, it becomes exactly the cumulative distribution. Moreover, this physical meaning applies to any particle shape besides round particles, and it is easily generalized to the rotationally averaged cumulative size distributions for non-rotationally symmetric cases, such as the cubic-shaped KCl powder particles shown in Fig.2a.

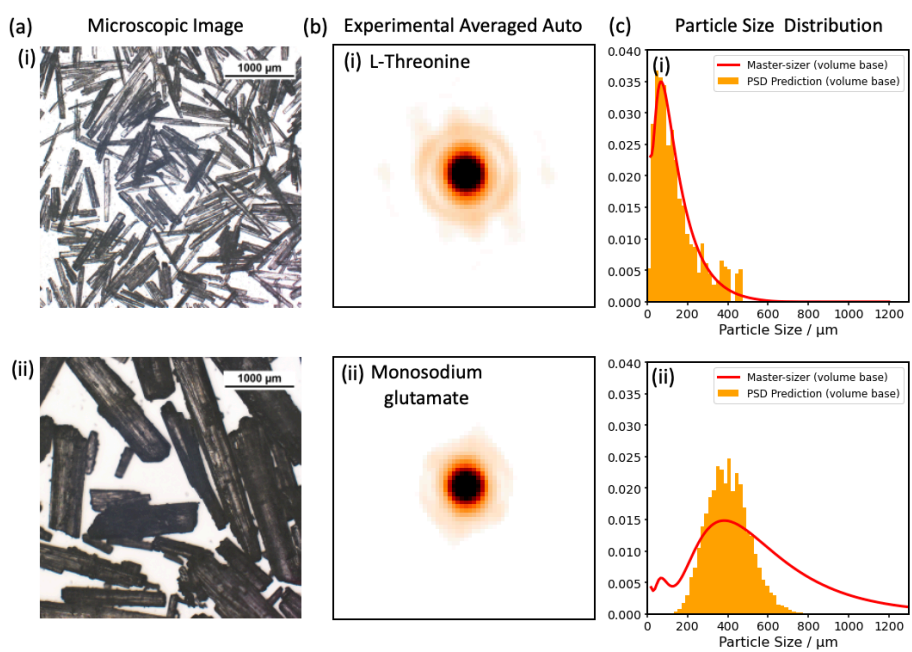

**Fig. S8 | Validation results for the needle shape powders L-threonine (i) and Monosodium glutamate, MSG (ii). (a)** In the microscopic images with the same scale bar for both (i) and (ii), MSG is clearly much bigger than L-threonine. **(b)** Measured averaged autocorrelations showing that MSG results in weaker side lobes compared to L-threonine. **(c)** Particle size distributions in volume base measured by our method (orange bar) and the Mastersizer (red line) serving as the ground truth.

To validate the performance for needle shape particles, we trained the pipeline with speckles from sieved KCl particles and tested it with L-threonine and Monosodium glutamate (MSG). The result is shown in Fig. S8. The microscope images in (a) show a visually clear size difference between these two powders, and verify that they are all needle-shaped. Part (b) shows averaged autocorrelations. According to. Fig. 2g in the main text, the first order side lobe is merged into the main lobe and is hard to resolve. L-threonine has a stronger second order side lobe than MSG. Higher order side lobes cannot be observed for either material. The particle size distributions transformed into volume base are plotted in part (c). The L-threonine's speckle prediction matches the Mastersizer result. The prediction for MSG has the same peak position as the ground truth but different widths. We may explain this width mismatch from two viewpoints. The first one is that the ground truth size distribution disperses more than 1000 μm (1 mm), which is out of the range in our training sets because KCl never forms particles as large. The second reason is that our model cannot work with the bimodal distribution very well, as we mention in the Discussion section of the main test and further discuss in Supplementary Section 6. This experiment confirms that, to a certain extent, our model can apply to different particle shapes without retraining the neural network.

**Details of the training process.** The learning rate for the generator **G** is fixed to be $10^{-2}$ to avoid overfitting. The batch size is set to be 4. The training loss function is the negative Pearson correlation coefficient (NPCC)[6,7]. If $k$ denotes the index in the training batch, and $(i,j)$ denotes the $(i,j)^{th}$ th pixel in a particular image, then the batch-wise NPCC loss is defined as:

$$L_{\text{NPCC}} = \sum_k \varepsilon_{\text{NPCC}}(f_k, \hat{f}_k)$$

$$\varepsilon_{\text{NPCC}}(f_k, \hat{f}_k) = -\frac{\sum_{i,j}(f_k(i,j) - \overline{f_k})(\hat{f}_k(i,j) - \overline{\hat{f}_k})}{\sqrt{\sum_{i,j}(f_k(i,j) - \overline{f_k})^2}\sqrt{\sum_{i,j}(\hat{f}_k(i,j) - \overline{\hat{f}_k})^2}} \quad (M23)$$

Here, ‾ denotes spatial averaging. The ideal minimum value of this loss function is -1. Our validation NPCC loss can reach -0.995 after 50 epochs of training.

The learning rate of the estimator **F** was set to be $2 \times 10^{-4}$ initially and halved whenever validation loss plateaued for 6 consecutive epochs. Batch size was set to 4. The training lasted for 100 epochs with mean absolute error (MAE) as the training loss function. For a 1D probability distribution, MAE is equivalent to the 1-Wasserstein distance[10]. The validation MAE loss for the synthetic data can reach 0.017 while it is 0.027 for the experimental validation data. This small mismatch is within our tolerance.

The computer used for training has Intel Xeon G6 CPU, 128 GB RAM, and dual Volta GPUs with 64 GB VRAM. Both training and test data were shuffled before each epoch. It took around 2 minutes to finish the whole training process.

## 5. Estimator results for the G-training dataset.

Fig. S8 plots the results for the dataset used to train the generator G. These data were disjoint from the training process of **F**. This plot is to cross-validate that **G** is not overfitting. If there is an overfitting effect, the performance of this dataset should be better than the test dataset. Compared to Fig.3, these two datasets have a similar performance from the second and the fifth columns, which indicates a well-trained generator.

From the first line result (50 μm – 106 μm) in this figure, we want to address that the prediction at the small size end has more deviation than the big size end. This phenomenon can also be interpreted by Equation (1), because the basis $\frac{\sin(ru)}{u}$ approaches 0 as $r$ approaches 0, making its weight harder to distinguish at small radii. Moreover, we can reach a smaller $r$ by modifying our optical system to generalize the interested size range in other applications. Based on the relationship $u = \frac{u'}{\lambda f_3}$, we can reduce the focal length of L3 $f_3$ to keep $ru$ at the same value but with a smaller $r$ and a larger $u$. In this way, we can push the lower bound even lower at the cost of compromising accuracy in the big size end of the predictions.

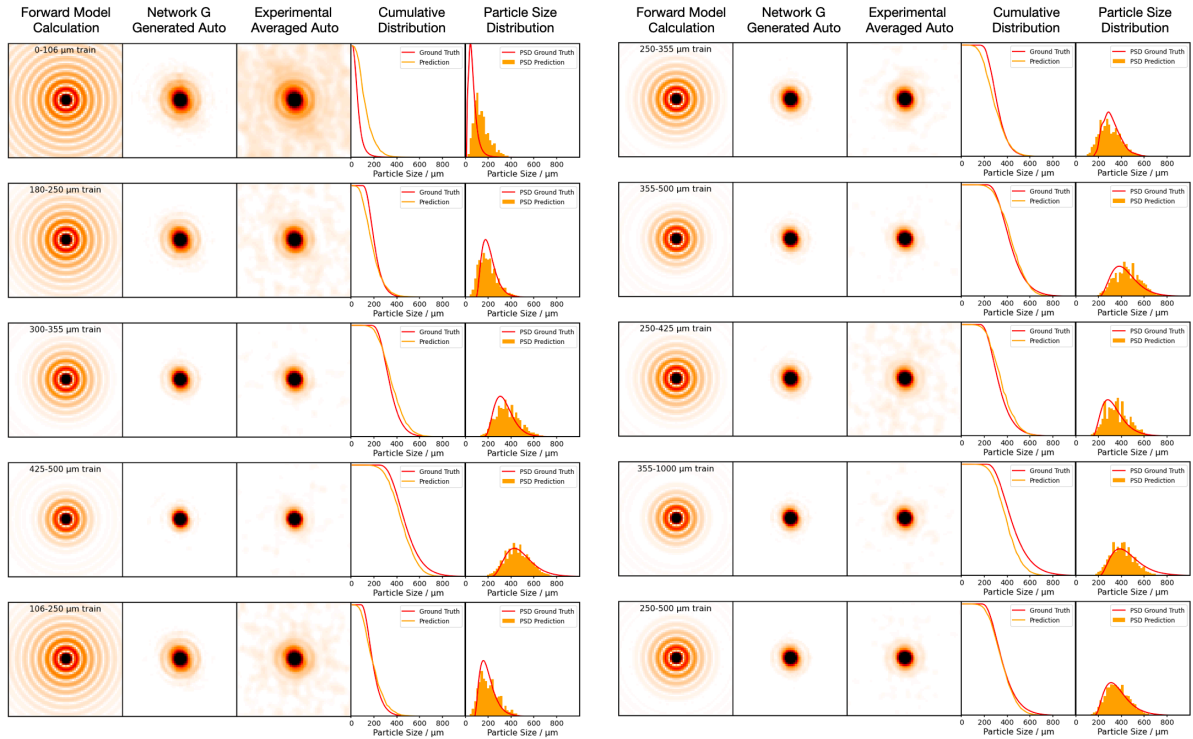

**Fig. S9 | Prediction results by the generator and estimator with the generator G training dataset.** Results of ten **G** training sample sets are plotted. These data were disjoint from the training process of the estimator **F**. The first and second columns show the output from the forward operator **H** and the generator **G**, respectively. The measured averaged autocorrelations are plotted in the third column. The prediction (marked as orange) from the estimator **F** and ground truth (marked as red) of the cumulative distributions and the corresponding PSDs are plotted together in columns 4 and 5.

## 6. Estimator stress test with a double peak PSD synthetic dataset.

We created a synthetic dataset with double-peak PSDs to test the estimator's generalization ability to different PSD shapes. Unfortunately, it fails to predict the double peaks PSDs, as shown in Fig. S10. The output is a single peak PSD located between the two peaks. The position is approximately influenced by the weight of these two peaks. This is partially because this double peak feature is not included in the training dataset, so it has not been learned as a prior to the PEACE pipeline. However, even if we train the estimator with a mixture of single and double-peak PSD datasets, it cannot provide a good estimation. We can explain this with the forward model. To resolve a single peak in the PSD, the estimator could combine the low-order lobes' intensity with the learned prior from the dataset to estimate. However, high-order lobes are required to distinguish double peaks in the PSD. However, the higher-order lobes will merge into the background fluctuations when the large particles exist, as shown in Fig. 2e and the second column in Fig. S10. We may be able to solve this issue by increasing the number of averaging frames to suppress the background fluctuation feature in the autocorrelation image, or by exploiting higher order correlations (cumulants). This problem is outside the scope of our present work, but interesting as a topic for future research.

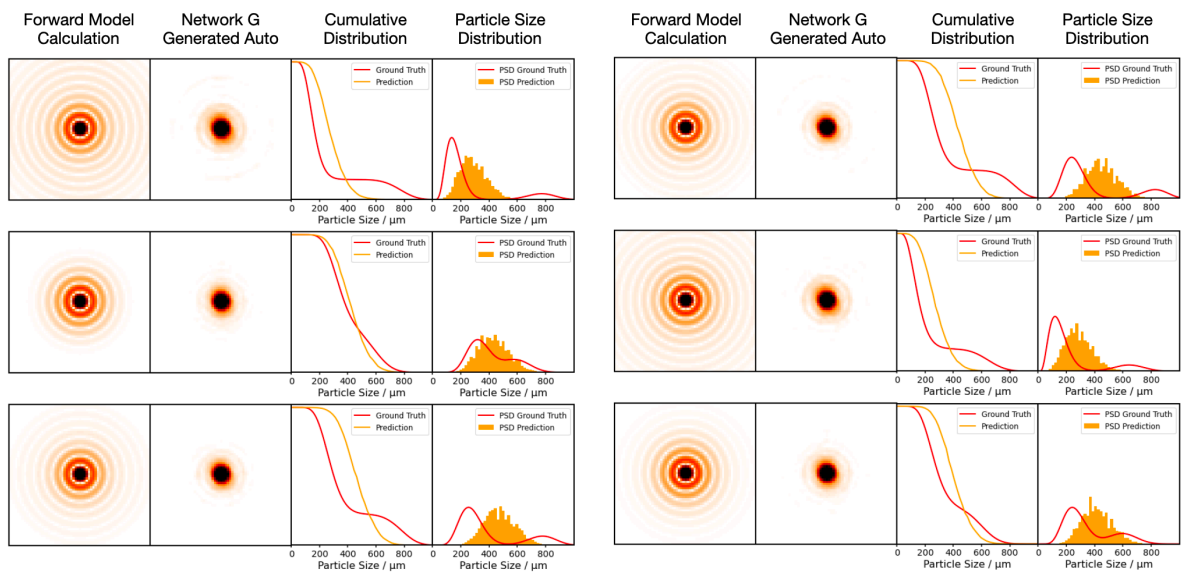

**Fig. S10 | The stress test results with a double peak PSD synthetic dataset.** The estimator fails to predict a double peak PSD. Instead, the estimation is a single peak PSD with the peak located between the two peaks.